\documentclass[prd,twocolumn,nofootinbib]{revtex4-2}

\usepackage{times}
\usepackage{amsmath}
\usepackage{amsfonts}
\usepackage{graphicx}
\usepackage{xspace}
\usepackage{hyperref}
\usepackage{comment}
\usepackage{dcolumn}
\usepackage{bm}

\def\nn{\nonumber}

\def\Oas{\mathcal{O}(\alpha_s)}

\DeclareMathOperator{\De}{d}
\newcommand{\de}{\De\!}

\newcommand{\cf}{C_{\text F}}
\newcommand{\ca}{C_{\text A}}

\newcommand{\sherpa}{S\protect\scalebox{0.8}{HERPA}\xspace}
\newcommand{\pythia}{P\protect\scalebox{0.8}{YTHIA}\xspace}
\newcommand{\herwig}{H\protect\scalebox{0.8}{ERWIG}\xspace}

\newcommand{\fastjet}{F\protect\scalebox{0.8}{AST}J\protect\scalebox{0.8}{ET}\xspace}
\newcommand{\rivet}{R\protect\scalebox{0.8}{IVET}\xspace}

\newcommand{\mgamcnlo}{M\protect\scalebox{0.8}{ADGRAPH}{5\_aMC@NLO}\xspace}

\newcommand{\softdrop}{\textsf{SoftDrop}}
\newcommand{\fjcontrib}{\textsf{fjcontrib}}

\newcommand{\lsim}{\lesssim}
\newcommand{\kt}{k_{t}}

\newcommand{\as}{\alpha_s}

\let\originalleft\left
\let\originalright\right
\renewcommand{\left}{\mathopen{}\mathclose\bgroup\originalleft}
\renewcommand{\right}{\aftergroup\egroup\originalright}

\newcommand{\zc}{z_{\rm cut}}

\newcommand{\zcut}{\ensuremath{z_{\text{cut}}}}

\def\beq{\begin{equation}}  
\def\eeq{\end{equation}}
\def\({\left(}
\def\){\right)}
\def\[{\left[}
\def\]{\right]}

\usepackage{color}
\definecolor{darkblue}{rgb}{0,0,0.5}
\definecolor{darkred}{rgb}{0.5,0,0}
\definecolor{darkgreen}{rgb}{0,0.5,0}
\definecolor{orange}{rgb}{1.0,0.49,0}

\begin{document}
\begin{minipage}[t]{0.95\textwidth}
    \raggedleft
         JLAB-THY-24-4244
\end{minipage}
\title{\boldmath Heavy Flavor Jet Substructure at Lepton Colliders}
\author{Prasanna K. Dhani~$^{(a)}$} \email{dhani@ific.uv.es}
\author{Oleh Fedkevych~$^{(b,c,d)}$} \email{ofedkevych@gsu.edu} 
\author{Andrea Ghira~$^{(e)}$}\email{andrea.ghira@ge.infn.it}
\affiliation{${}^{a}$ Instituto de F\'{\i}sica Corpuscular, Universitat de Val\`{e}ncia -- Consejo Superior de Investigaciones Cient\'{\i}ficas, Parc Cient\'{\i}fic, E-46980 Paterna, Valencia, Spain. \\
${}^{b}$ Physics and Astronomy Department, Georgia State University, Atlanta, GA 30303, USA. \\
${}^{c}$ Center for Frontiers in Nuclear Science, Stony Brook University, Stony Brook, NY 11794, USA.\\
${}^{d}$ Jefferson Lab, Newport News, Virginia 23606, USA.\\
${}^{e}$ Dipartimento di Fisica, Universit\`a di Genova and INFN, Sezione di Genova,\\ Via Dodecaneso 33, 16146, Genoa, Italy.
}

\date{\today}

\begin{abstract}
We provide a detailed analysis of event-shape observables, namely the energy correlation function and jet angularity, for heavy-flavor jets produced in electron-positron collisions, focusing on quantum chromodynamics (QCD) interactions. Using modern jet substructure techniques, we investigate the dead-cone effect, where QCD radiation is suppressed around a heavy quark within an angle proportional to its mass. Our analysis achieves next-to-leading logarithmic accuracy, combined with partial fixed-order contributions, to improve the description of the transition near the dead-cone threshold. To ensure a comprehensive perspective, we compare our analytical results with predictions from the \pythia, \herwig, and \sherpa Monte Carlo simulations at past and future lepton colliders.
\end{abstract}

\maketitle

\section{Introduction}
\label{sec:intro}
Abundant production of jets--collimated beams of hadrons--in high-energy collider experiments like those at the Large Hadron Collider (LHC) provides a unique opportunity to test the Standard Model (SM)~\cite{Britzger:2017maj, CMS:2013vbb, ATLAS:2017qir, ATLAS:2021qnl, CMS:2014qtp, AbdulKhalek:2020jut, Harland-Lang:2017ytb, Pumplin:2009nk, Watt:2013oha, CMS:2021iwu, ALICE:2021njq} of particle physics and search for potential new physics beyond it~\cite{Soper:2010xk, Godbole:2014cfa, Chen:2014dma, Adams:2015hiv}. 
To achieve these goals, jet substructure techniques have become essential, enabling precise tests of Quantum Chromodynamics (QCD), high-accuracy measurements of the strong coupling $\alpha_{s} (\equiv g_s^2/4\pi)$, and effective jet tagging, among other applications (for a detailed overview, see~\cite{Marzani:2019hun}). 
An important and widely studied class of jet substructure observables are energy correlation functions (ECFs)~\cite{Larkoski:2013eya,Moult:2016cvt} and jet angularities~\cite{Larkoski:2014pca, Berger:2003iw, Almeida:2008yp}.

Achieving an accurate theory description of jets and their intricate structure requires a combination of fixed-order calculations and resummation techniques within perturbative QCD framework. 
Fixed-order calculations are involved due to the presence of multiple scales like jet transverse momentum, jet radius, quark masses, \textit{etc}., and they often fall short, as they tend to be dominated by large logarithmic contributions as a result of the hierarchy between these scales. 
To make predictions reliable over entire phase space regions, these calculations need to be supplemented with resummed contributions. 
Nevertheless, the resummation of jet substructure observables is non-trivial. 
The presence of phase space boundaries can introduce non-global effects~\cite{Dasgupta:2001sh} and the algorithmic nature of defining jets complicates achieving all-order factorisation.

Over the past decade, these challenges have been tackled by the theory community, leading to a deeper understanding of jet substructure through extensive QCD studies (see, for example, \cite{Marzani:2019hun} and references therein). 
In this context, grooming techniques have emerged as powerful tools. 
These algorithms, designed to clean a jet by removing contributions from wide-angle soft radiation, have significantly enhanced our ability to describe jet physics using perturbation theory. 
By reducing the impact of non-perturbative effects like hadronization and the underlying event, grooming techniques simplify the analytical structure of resummed results. 
Notably, the \mbox{\softdrop~algorithm~\cite{Larkoski:2014wba}} helps eliminate the logarithmic enhancements from wide-angle soft gluons, including the complex non-global logarithms. 
Thanks to these properties, calculations for observables using the \softdrop~have now achieved next-to-next-to-leading logarithmic (NNLL) accuracy and beyond~\cite{Frye:2016aiz,Kardos:2020gty}.

In recent years, precision studies of jets with heavy flavors like charm ($c$) and bottom ($b$) quarks have gained momentum in the jet substructure community~\cite{Cunqueiro:2018jbh, Lee:2019lge, Craft:2022kdo, Apolinario:2022vzg, Andres:2023ymw, Caletti:2023spr, Ghira:2023bxr, Gaggero:2022hmv,Ghira:2024nkk, Wang:2023eer, Zhang:2023jpe, Dhani:2024gtx, Jiang:2024qno, Dainese:2024wix, vonKuk:2024uxe, Aglietti:2024zhg}.
However, at the moment,  only a handful of measurements of jet substructure observables for jets seeded by heavy quarks is available~\cite{CMS:2024kzm, ALICE:2022phr, CMS:2024gds, CMS:2024kzm}.
These considerations play a key role for Higgs boson studies and offer insights into the heavy-quark content of the proton by examining heavy-flavor jets recoiling against electroweak bosons. 
In particular, the recent development of flavor-jet algorithms~\cite{Banfi:2006hf,Caletti:2022hnc,Czakon:2022wam,Gauld:2022lem,Caola:2023wpj} have opened the door to a novel yet-unexplored flavor-jet substructure program at high-energy colliders.
From a theoretical view point, the mass of the heavy quarks sets a perturbative scale for the running coupling and at the same time acts as a regulator of the collinear divergences. Experimentally, the relatively long lifetimes of $B$- and $D$-hadrons cause their decay to occur away from the primary interaction point. 
Dedicated $b$- and $c$- tagging techniques that capitalize on this property to identify $B$ and $D$ hadrons or $b$ and $c$ jets are widely used in collider experiments~\cite{ATLAS:2017bcq, ATLAS:2018nnq}.

One of the effects that influences the substructure of heavy flavored jets is the dead-cone effect~\cite{Dokshitzer:1991fd,Dokshitzer:1995ev, Ellis:1996mzs}. 
This phenomenon results in a suppression of collinear radiation near a heavy quark, altering the radiation pattern compared to light quarks. 
The first direct observation of this effect was recently reported by the ALICE collaboration~\cite{ALICE:2021aqk} (for indirect evidences, see Refs.~\cite{DELPHI:1992pnf, OPAL:1994cct,  OPAL:1995rqo, SLD:1999cuj, DELPHI:2000edu, ALEPH:2001pfo, ATLAS:2013uet}). 
From theory side, the case of ECFs involving heavy flavors produced in $e^+e^-$ collisions was considered in Ref.~\cite{Lee:2019lge} within the framework of soft-collinear effective theory~\cite{Bauer:2000ew,Bauer:2000yr,Bauer:2001yt,Bauer:2002nz}. 
Building on this, our recent work~\cite{Dhani:2024gtx} presented a comprehensive analysis of widely considered jet substructure observables such as ECFs and jet angularities specifically for flavored $b$-jets produced in $pp$ collisions. 
In this study, we delve into the implications of our findings for the Large Electron-Positron (LEP) collider data and look ahead to potential applications at future lepton colliders. Recent studies on the construction of an observable-independent resummation framework to account for heavy quark mass effects in $e^+e^-$ collisions can be found in Refs.~\cite{Ghira:2023bxr, Aglietti:2024zhg}. 

Precision measurements at lepton colliders have been pivotal in advancing our understanding of the SM of particle physics. In particular, they have provided deep insights into its renormalizability and the associated radiative corrections, with key contributions from measurements at the $Z$-boson pole at LEP~\cite{DELPHI:1992qrr, L3:1992btq,OPAL:1991uui,OPAL:1993pnw}.
As we look to the future, upcoming lepton colliders such as the International Linear Collider (ILC)~\cite{LCCPhysicsWorkingGroup:2019fvj}, the Circular Electron Positron Collider (CEPC)~\cite{CEPCStudyGroup:2018ghi}, and the Future Circular Collider with $e^+e^-$ (FCC-ee)~\cite{FCC:2018evy} promise to push the boundaries of precision even further. 
With luminosities around the $Z$-boson pole vastly surpassing those achieved at LEP, these colliders will open up new possibilities for exploration. 
Specifically, the ILC in its GigaZ operation will achieve luminosities 500 times greater than the total integrated luminosity at LEP, while the CEPC and FCC-ee could reach luminosity enhancements by factors of $5 \times 10^5$ to $8 \times 10^5$~\cite{Belloni:2022due}. 
This immense leap forward will dramatically reduce statistical uncertainties, allowing for measurements of the properties of $Z$ boson and other fundamental particles including the Higgs boson in the SM with an unprecedented level of precision and unlocking new frontiers in the quest to understand the Universe at its most fundamental level. 

To the best of our knowledge, only one study has measured jet substructure in $e^+e^-$ collisions~\cite{Chen:2021uws}.
These collisions offer the cleanest environment for jet production, free from multi-parton interactions (MPI), initial-state QCD radiation, and parton distribution function uncertainties. The absence of soft radiation from proton remnants uniquely allows the study of hadronization effects on jet substructure without interference from MPI $-$ unlike in $pp$ collisions, where these contributions overlap. 
Experimental investigations of $e^+e^-$ collisions can not only test the accuracy of theory predictions but also provide valuable data to refine non-perturbative models implemented in general-purpose Monte Carlo (MC) event generators such as \pythia~\cite{Sjostrand:1985vv, Sjostrand:1987su, Sjostrand:2004pf, Andersson:1983ia, Sjostrand:1984ic}, \herwig~\cite{Bahr:2008dy,Gieseke:2016fpz,Webber:1983if,Kupco:1998fx}, and \sherpa~\cite{Winter:2003tt} as highlighted in~\cite{Aguilar:2021sfa,Reichelt:2021svh}.

The article is structured as follows: in Section~\ref{sec:observables} we define our observables and briefly discuss the grooming procedure we follow. 
We then dive deep into their perturbative analysis at the first non-trivial order in Section~\ref{sec:fo_analysis}, before moving on to all-order analysis in Section~\ref{sec:resum_analysis}. 
In Section~\ref{sec:numerics} we compare our predictions with general-purpose MC event generators before concluding with our final thoughts in Section~\ref{sec:conclusions}.
The detailed analytical results at $\mathcal{O}(\as)$ are left to Appendix~\ref{app: fixed order results} and the resummed cumulative distributions for light-quark jets are given in Appendix~\ref{sec:resummed-radiators-light-quarks}.

\section{Definition of the observables}
\label{sec:observables}
This section highlights jet substructure observables for studying heavy flavor jets, focusing on ECFs~\cite{Larkoski:2013eya,Moult:2016cvt} and jet angularities~\cite{Larkoski:2014pca, Berger:2003iw, Almeida:2008yp}.
Our analysis examines QCD matrix elements in the quasi-collinear limit~\cite{Catani:2002hc,Catani:2000ef}  relevant for jets initiated by massive quarks (for the extension of quasi-collinear factorisation at $\mathcal{O}(\as^2)$, see Refs.~\cite{Dhani:2023uxu,Craft:2023aew}). 
Within this limit, both the transverse momentum of emitted radiation, $k_t$, and the heavy quark mass, $m$, are considered small compared to the hard scale $Q$, but their ratio $k_t/m$ is kept fixed. 
The additional mass scale introduces large-logarithmic contributions depending on the ratio $m^2 / Q^2$, which can be systematically resummed using perturbative QCD, as detailed in Section~\ref{sec:resum_analysis}.

We now introduce the jet substructure observables and setup used in our study. We focus on $e^+e^-$ collisions in the back-to-back (di-jet) limit at LEP and ILC, with center-of-mass energies of 91 GeV and 2 TeV, respectively. 
Unlike $pp$ collisions, we do not analyze small-radius jets but instead consider contributions from the ``upper'' and ``lower'' hemispheres defined by the final-state partonic configuration. 
From an MC perspective, this corresponds to clustering events into two exclusive jets using the $e^+e^-$ generalization of the $k_t$-family clustering algorithm~\cite{Catani:1991hj,Dokshitzer:1997in,Cacciari:2008gp}, implemented with the standard $E$-scheme in the \fastjet~\cite{Cacciari:2011ma} library. Here, the observables are defined using energy fractions $z_i= 2 E_i/\sqrt{s}$, where $E_i$ is the energy of the $i^{\rm th}$ particle in the jet, and $\sqrt{s}\equiv Q$ is the center-of-mass energy of the $e^+e^-$ system.

In light of the study conducted in \cite{Dhani:2024gtx}, in this work we only focus on the plain definition of ECFs and jet angularities. We begin by introducing the ECF which, in our case, is given by
\begin{align}
\label{eq:e2}
e_2^\alpha= \sum_{\mathcal{H}}\sum_{i,j\in\mathcal{H}, i<j}z_i z_j \left[2(1-\cos \theta_{ij}) \right]^\frac{\alpha}{2}\Theta\left((\vec{p}_i\cdot\vec{n})(\vec{p}_j\cdot\vec{n})\right),
\end{align}
where the first sum runs over two hemispheres $\mathcal{H}$ and the step function ensures that particles $i$ and $j$ remain in the same hemisphere defined by the jet axis vector $\vec n$. To maintain infrared and collinear (IRC) safety, we require $\alpha > 0$.
%
Let us now consider a closely related observable, commonly known as jet angularity, which is defined as follows
\begin{align}
\label{eq:lambda}
\lambda^\alpha= \sum_\mathcal{H} \sum_{i\in\mathcal{H}}&z_i  \left[2(1-\cos \theta_{i}) \right]^\frac{\alpha}{2},
\end{align}
where $\theta_i$ is the angle between $i^{\rm th}$ particle and jet axis defined with the Winner-Takes-All recombination scheme to avoid contribution from recoil effects~\cite{Banfi:2004yd, Larkoski:2013eya, Larkoski:2014uqa}.
%

We also analyze the groomed variants of these observables, obtained by first applying the \softdrop~\cite{Larkoski:2014wba} to each of the two hemispheres. The same expressions as above in Eqs.~(\ref{eq:e2}) and (\ref{eq:lambda}) are then computed on the groomed hemispheres. The \softdrop~procedure involves reclustering the jet, traversing its angular-ordered branching history, and sequentially removing softer branches until the condition
\begin{equation}
\label{eq:SDee}
\frac{\text{min}\{E_i,E_j\}}{E_i+E_j}> \zcut \left[2(1-\cos{\theta_{ij}})\right]^\frac{\beta}{2}
\end{equation}
is satisfied. Here, $i$ and $j$ denote the branches at a given step in the clustering process. Both fixed order and resummed calculations are performed for $\beta=0$ which corresponds to the modified Mass Drop Tagger (mMDT)~\cite{Dasgupta:2013ihk}, and for $z_{\rm cut}\ll 1$ so that power corrections in $z_{\rm cut}$ are systematically neglected.

The choice of mMDT is driven by its ability to isolate the effects of the dead-cone phenomenon on jet substructure observables. By focusing on collinear dynamics, mMDT effectively suppresses contributions from soft emissions, allowing us to better understand the impact of massive quarks on jet substructure. 
For our MC analysis, we utilize the mMDT implementation provided by the \fjcontrib~plugin for the \fastjet~library.

\section{Fixed order analysis}
\label{sec:fo_analysis}

In this section, we first analyse the observables defined in Eqs.~(\ref{eq:e2}) and (\ref{eq:lambda}) using the fixed-order perturbation theory. 
In particular, we will calculate the cumulative distribution for the ECF and the jet angularity and therefore define
\begin{equation}
\label{eq:cum-dist}
\Sigma_V(v) = \frac{1}{\sigma_0}\int_0^v {\rm d} v' \frac{{\rm d}\sigma_V}{{\rm d} v'},
\end{equation}
i.e.~the probability that the observable $V$ is smaller than a certain value $v$. In Eq.~(\ref{eq:cum-dist}), the normalization, $\sigma_0$, denotes the total Born cross section. We consider the $\mathcal{O}(\as)$ contribution to $\Sigma_V(v)$ i.e. $\Sigma_V^{(1)}(v)$.
This corresponds to the one-gluon ($g$) emission, plus $\mathcal{O}(\as)$ virtual contributions to the leading order (LO) process \mbox{$e^+ + e^-\to b + \bar{b}$.}
As discussed earlier, QCD scattering amplitudes with massive quarks factorise in the quasi-collinear limit. 
In this approximation, the transverse momentum of the emitted radiation $k_t$ and the mass of the heavy quark $m$ are assumed to be of the same order and small compared to the hard scale $\sqrt{s}$ of the process. 
Using the standard Sudakov decomposition, we obtain the double quasi-collinear factorisation at tree level as follows
\begin{equation}
\label{eq:quasi-collinear-fact}
|\mathcal{M}|^2\simeq \frac{8\pi \as z(1-z)}{k_t^2+z^2 m^2} P_{gb}(z,k_t^2) |\mathcal{M}_0|^2,
\end{equation}
where $\mathcal{M}$ denotes the original scattering amplitude and $\mathcal{M}_0$ denotes the reduced amplitude with one less external parton. In Eq.~(\ref{eq:quasi-collinear-fact}), $z$ is the fraction of the momentum transferred in the splitting process $b \rightarrow g+ b$, and $P_{gb}$ is the LO time-like massive splitting function~\cite{Catani:2002hc}
\begin{align}
P_{gb}(z,\kt^2)= \cf \left[\frac{1+(1-z)^2}{z}-\frac{2 m^2 z(1-z)}{\kt^2+z^2 m^2}\right].
\end{align}

To calculate the $\mathcal{O}(\as)$ contribution $\Sigma_V^{(1)}$, we integrate Eq.~(\ref{eq:quasi-collinear-fact}) together with a $\Theta$ function that forces the observable to be smaller than a certain value $v$. 
In the resummation language, this corresponds to the first-order approximation of minus the radiator, $-\mathcal{R}^{(\text{f.o.})}_V$. 
We therefore evaluate 
\begin{align}
\label{eq:sudakov-fo-ee}
\mathcal{R}^{(\text{f.o.})}_V(v,\xi)=& -\frac{\as}{2\pi} \int^{\theta_{\rm max}^2}_0 \frac{{\rm d}\theta^2}{\theta^2+4\xi} \int^1_0 {\rm d} z\,P_{gb}\left(z,k_t^2\right)
\nn\\
&\times\Big[ \Theta(v-V(z,\theta^2)) -1\Big],
\end{align}
where  $\xi = m^2/s$, and $k_t^2\simeq sz^2\theta^2/4$ with $\theta$ the angle between the emitting \mbox{$b$-quark} and the gluon. 
The upper limit $\theta_{\rm max}^2$ is irrelevant for our analysis as we focus on the logarithmic integration structure, on $\log\xi$ and $\log v$. 
The contribution $-1$ takes into account virtual corrections, while $V(z,\theta^2)$ parameterises the observable in the quasi-collinear limit. 

We begin with the angularity, for which we can readily obtain a close analytical expression by inserting the value of the observable in the quasi-collinear limit, $V=z\theta^{\alpha}$, into Eq.~(\ref{eq:sudakov-fo-ee}). 
The calculation for the ECF \mbox{$e_2^\alpha$ $(V = z(1-z)\theta^{\alpha})$} is more involved. 
The calculation of radiators in the quasi-colinear limit is performed neglecting power corrections in both the mass $m$ and in the substructure observable $v$, while keeping the ratio \mbox{$x=\xi/\bar{v}^\frac{2}{\alpha}$} fixed with $\bar v= v/ 2^\alpha$.
For complete analytical results at $\Oas$, we refer to the Appendix~\ref{app: fixed order results} (see also, Ref.~\cite{Dhani:2024gtx} for further details).

We now examine radiator's behaviour for small and large values of $x$. 
The former is the limit where the quark mass is much smaller than the observable, and therefore we expect to recover the massless result. 
In the large-$x$ limit, we expect to probe the dead-cone region instead.
For angularity and ECF we find
\begin{subequations} 
\label{eq:lambda-e2-alpha-x}
\begin{align} 
  \label{eq:lambda-e2-alpha-smallx}
    \mathcal{R}^{(\text{f.o.})}_{V}(v,\xi)\overset{x\ll 1}{\simeq}&\frac{\as \cf}{\pi}\left(\frac{1}{\alpha}\log ^2 \bar{v}+\frac{3}{2 \alpha}\log \bar{v}+ \mathcal{K}_V\right),
\\
\label{eq:lambda-e2-alpha-largex}
  \mathcal{R}^{(\text{f.o.})}_{V}(v,\xi)
  \overset{x\gg 1}{\simeq}&\frac{\as \cf}{\pi} \Bigg[\log \xi\, \log \bar{v}-\frac{\alpha}{4}\log^2 \xi
  \nn\\
  &+ \left(\frac{3}{4}-\frac{\alpha}{2}\right)\log{\xi}+\log \bar{v}-\frac{\alpha\pi^2}{12}+1\Bigg],
\end{align}
\end{subequations}
where $\mathcal{K}_{\lambda^{\alpha}} = \frac{7}{4\alpha}$ and $\mathcal{K}_{e_2^{\alpha}} = \frac{1}{\alpha}\left(3-\frac{\pi^2}{3}\right)$.
We observe that Eqs.~(\ref{eq:lambda-e2-alpha-smallx}) exhibits a double logarithmic behaviour due to soft and collinear emissions. 
Conversely, in Eq.~(\ref{eq:lambda-e2-alpha-largex}), the double logarithms in $v$ are canceled, resulting in double logarithms of $\xi$. 
The ECF starts to differ from its angularity counterparts for massless quarks beyond the next-to-leading logarithmic (NLL) accuracy. 
In contrast, within the dead-cone region, both the ECF and the angularity exhibit the same behavior.
Next, we calculate the $\Oas$ contribution for mMDT groomed jets. 
At this order, the jet consists of a massive quark and a gluon, and the groomed-jet radiator (denoted by a bar) reads
\begin{align}
\label{eq:sudakov-fo-groomed-ee}
\bar{\mathcal{R}}^{(\text{f.o.})}_V(v,\xi)=& \frac{\as}{2\pi} \int^{\theta_{\rm max}^2}_0 \frac{{\rm d}\theta^2}{\theta^2+4\xi} \int^{1-\zc}_{\zc} {\rm d} z\,P_{gb}\left(z,k_t^2\right) 
\nn\\
\times&\Theta\left(V(z,\theta^2)-v\right).
\end{align}
We explicitly calculate for $\lambda^\alpha$ and note that the procedure is identical for $e_2^{\alpha}$. In the groomed region, $v<z_{\rm cut}$, and in the limit $z_{\rm cut}\ll 1, x\zc^{\frac{2}{\alpha}}\ll 1$, we have
\begin{align}
\label{eq:lambda-alpha-groomed-small-zcut}
 \bar{\mathcal{R}}^{\text{(f.o.)}}_{{\lambda}^\alpha}\left(v,\xi \right) &\simeq \mathcal{R}^{\text{(f.o.)}}_{{\lambda}^\alpha}\left(v,\xi\right) + \frac{\as \cf}{\pi}\Bigg[- \frac{1}{\alpha}\log^2 \frac{\bar{v}}{z_{\rm cut}}\Bigg].
\end{align}
The $\Oas$ constant term, independent of $v, \xi,$ and $\zc$, remains unchanged between the ungroomed and groomed cases due to its origin in hard-collinear radiation, which mMDT retains up to power corrections in $\zc$. 
This consistency, also observed for the ECF in the current study, aligns with physical expectations. 
We leverage this simplification in Section~\ref{sec:resum_analysis} to address transition-point corrections in the NLL resummation. 
For further details we recommend the reader to the Ref.~\cite{Dhani:2024gtx}.
\section{All order analysis}
\label{sec:resum_analysis}
We now present the calculation to all perturbative orders, which enables the resummation of large logarithmic contributions due to the various scale hierarchies. 
We start with ungroomed jets and then move on to the mMDT groomed jets. 
We present our calculation that resums the logarithms of $v$ and the logarithms of $\xi$ in each of the regions defined by the hierarchy between $v$ and $\xi$ with NLL accuracy. 
This leads to a simplification and we only need to calculate one radiator, \mbox{i.e. $\mathcal{R}_V(v,\xi) \xrightarrow{\text{NLL}} R(v,\xi)$.} 
%

%
The radiator is closely related to the fixed-order calculation presented in Eq.~(\ref{eq:sudakov-fo-ee}). It can be expressed as
\begin{align} 
\label{eq:radiator-ungroomed}
R_b(v,\xi)=&\int^{1}_0 {\rm d} z \int^{1}_{\xi} \frac{{\rm d} \theta^2}{\theta^2} {P}_{gb}\left(z,\bar{k}_t^2-z^2 m^2\right)
\nn\\
&\times\frac{\as^{\text{CMW}}(\bar{k}_t^2)}{2\pi} \Theta\left(z\theta^{\alpha}-\bar{v}\right),
\end{align}
where $\bar{k}_t^2 = sz^2\theta^2$. The running coupling is considered in the Catani-Marchesini-Webber (CMW) scheme \cite{Catani:1990rr}.
To fully account for mass effects, the strong coupling is treated in the decoupling scheme: $\as(\bar k_t^2)= \as^{(5)}(\bar k_t^2) \Theta(\bar k_t^2-m^2)+\as^{(4)}(\bar k_t^2) \Theta(m^2-\bar 
k_t^2)$, and the result is expressed using the five-flavor coupling evaluated at the hard scale, \mbox{$\as^{(5)}(s)\equiv\as$},
\begin{subequations}
	
	\begin{align}
	\label{eq: dec_a}
		\as^{(5)}(\bar k_t^2)=& \frac{\as}{1-\nu^{(5)}}\left(1-\as \frac{\beta_1^{(5)}}{\beta_0^{(5)}}\frac{\log{(1-\nu^{(5)})}}{1-\nu^{(5)}}\right),\\
		\label{eq: dec_b}
		\as^{(4)}(\bar k_t^2)=& \frac{\as}{1-\nu^{(4)}-\delta_{54}} \nonumber \\
		&\times\left(1-\as \frac{\beta_1^{(4)}}{\beta_0^{(4)}}\frac{\log{\left(1-\nu^{(4)}-\delta_{54}\right)}}{1-\nu^{(4)}-\delta_{54}}\right),
	\end{align}
\end{subequations}
where
\begin{align}\label{eq:nu-def}
	\nu^{(n_f)}&= \as \beta_0^{(n_f)}
	\log{\left(\frac{s}{\bar{k}_t^2}\right)}, \qquad n_f=4,5 \nonumber,\\
	\delta_{54}&= \as \left(\beta_0^{(5)}-\beta_0^{(4)}\right)\log{\left(\frac{1}{\xi}\right)}.
\end{align}
In the above expressions, we introduced the one- and two-loop coefficients of the QCD $\beta$-function.
%
%
The quasi-collinear radiator with grooming $\bar{R}_b(v,\xi)$ is a generalization of Eq.~(\ref{eq:radiator-ungroomed}). Its expression is given by
\begin{align} 
\label{eq:radiator-groomed}
\bar{R}_b(v,\xi)=&\int^{1}_0 {\rm d} z \int^{1}_{\xi} \frac{{\rm d} \theta^2}{\theta^2}P_{gb}\left(z,\bar{k}_t^2- z^2 m^2\right)
\nn\\
&\times\frac{\as^{\text{CMW}}(\bar{k}_t^2)}{2\pi} \Theta\left(z\theta^{\alpha}-\bar{v}\right)\Theta(z-\zc).
\end{align}
Both ungroomed and groomed radiators in Eqs.~(\ref{eq:radiator-ungroomed}) and (\ref{eq:radiator-groomed}) are computed exploiting Lund diagrams for heavy flavors, introduced in \cite{Ghira:2023bxr}. 
We note that they are continuous in \mbox{$\bar{v}= \xi^\frac{\alpha}{2}$}, but their derivatives are not. This leads to discontinuous differential distributions at the transition. We can employ fixed-order calculations, particularly the non-logarithmic contributions from Eqs.~(\ref{eq:sudakov-fo-ee}) and (\ref{eq:sudakov-fo-groomed-ee}), to smooth this behavior and obtain a better theoretical description of the region near the dead-cone.

Up to NLL accuracy, the all-order cumulative distribution in the ungroomed case reads \cite{Catani:1992ua, Banfi:2004yd}
\begin{align}\label{eq:sigma_b_w_corr}
\Sigma_V^b(v,\xi)=& \frac{e^{-2\gamma_{\text{E}} R_b'(v,\xi)}}{\Gamma\left(1+ 2R_b'(v,\xi)\right)} e^{-2R_b(v,\xi)}\nonumber \\
 &\times \exp \left[-2\left( \mathcal{R}^{(\text{f.o.})}_V(v,\xi)- R_b^\text{(f.c.)}(v,\xi)\right)\right],
\end{align}
where $R'_b = \partial R_b / \partial L , L=- \log v$. In the above expression, the factor of 2 arises from considering the di-jet configuration, which involves summing over both hemispheres.  In Eq.~(\ref{eq:sigma_b_w_corr}), $R_b^\text{(f.c.)}$ denotes the fixed coupling expansion of Eq.~(\ref{eq:radiator-ungroomed}) which needs to be subtracted from $\mathcal{R}^{(\text{f.o.})}_V(v,\xi)$ to avoid double-counting of large logarithmic contributions. Analogously, the all-order groomed cumulative distribution is given by
\begin{align}
\bar{\Sigma}_V^b(v,\xi)=& \frac{e^{-2\gamma_{\text{E}} \bar R_b'(v,\xi)}}{\Gamma\left(1+ 2\bar R_b'(v,\xi)\right)} e^{-2\bar R_b(v,\xi)}\nonumber \\
&\times\exp \left[-2\left( \mathcal{R}^{(\text{f.o.})}_V(v,\xi)- R_b^\text{(f.c.)}(v,\xi)\right)\right].
\end{align}
We emphasize that the fixed-order contributions are identical for both the ungroomed and groomed cases, as the $\mathcal{O}(\alpha_s)$ calculation of the groomed radiator yields the same constant terms, up to power corrections in $\zc$. The corresponding resummed cumulative distributions for light-quark jets which we use for making numerical predictions in Section~\ref{sec:numerics} are presented in Appendix~\ref{sec:resummed-radiators-light-quarks}.
\section{Numerical predictions}
\label{sec:numerics}
Using the fixed-order and all-order analyses discussed in Sections~\ref{sec:fo_analysis} and \ref{sec:resum_analysis}, we predict outcomes for LEP and future lepton collider energies.
Specifically, we analyze the ratio of cumulative distributions for $b$-jets to light quark jets, utilizing massless ungroomed and groomed radiators derived by setting $\xi=0$ in Eqs.~(\ref{eq:radiator-ungroomed}) and (\ref{eq:radiator-groomed}). 
Our results are compared with parton-level simulations obtained using general-purpose MC event generators like \pythia\cite{Bierlich:2022pfr,Sjostrand:2006za}, \herwig~\cite{Bellm:2019zci,Corcella:2000bw}, and \sherpa~\cite{Sherpa:2019gpd,Gleisberg:2008ta}. 
The classical leading-log accurate dead-cone angle, \mbox{$\theta_D = 2 m/ \sqrt{s}$}, is determined from the lower bound of $\theta^2$ integrations in Eqs.~(\ref{eq:radiator-ungroomed}) and (\ref{eq:radiator-groomed}), and increases as the heavy quark's energy decreases due to subsequent emissions (for additional details see Ref.~\cite{Aglietti:2024zhg}). 
The dead-cone threshold is shown as a vertical dashed black line in our plots.

All resummed contributions indicated by solid black lines are supplemented with an uncertainty band. This band is calculated by rescaling the argument of the strong coupling by $x_R$ and rescaling the observable by $x_v$. Appropriate counter terms are added to maintain NLL accuracy. To estimate the theoretical uncertainty in the region around the dead-cone transition, we vary both $x_R$ and $x_v$ within specific ranges. We consider the cases where $x_R=1$ and $\frac{1}{2}\leq x_v\leq 2$, and where $\frac{1}{2}\leq x_R\leq 2$ and $x_v=1$. We obtain the uncertainty band by choosing the lowest and highest envelope out of five possible combinations, with the central curve being $(x_R=1, x_v=1)$.

\begin{figure}
\begin{center}
\includegraphics[width=0.53\textwidth]{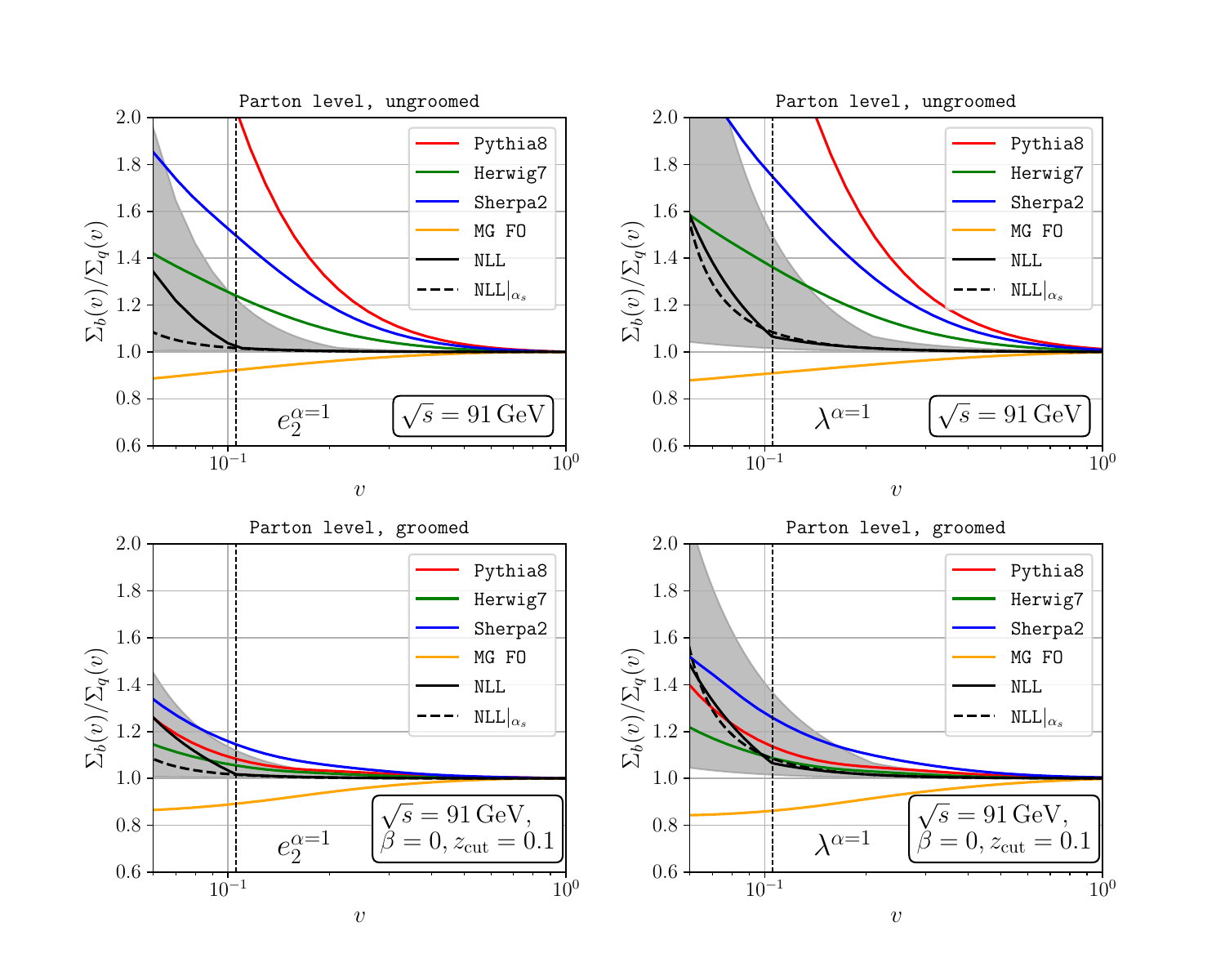}
\caption{The ratio $\Sigma^b / \Sigma^q$ at the parton level at LEP center-of-mass collision energy is presented. Our NLL predictions are in solid black, with their truncation at $\Oas$ indicated by dashed black lines and the full real matrix element at NLO in yellow obtained using \mgamcnlo. The predictions from \pythia, \herwig, and \sherpa MC simulations are in color. The vertical dashed black line marks the dead-cone transition threshold \mbox{$\theta_D = 2m_b / \sqrt{s} \approx 0.11$.}  All MC simulations are performed at LO+PS accuracy.}
\label{fig:LEP-all_pl}
\end{center}
\end{figure}

Let us discuss the LEP collision energy \mbox{$\sqrt{s} = 91$ GeV}. 
We set the $b$-quark mass $m = 4.8$ GeV which implies $\theta_D \approx 0.11$. 
The solid black line in Fig.~\ref{fig:LEP-all_pl} shows the resummed predictions for the ratio $\Sigma^b / \Sigma^q$ of cumulative distributions.
For observable values larger than $\theta_D$, the ratio remains mostly equal to unity. 
However, as $v$ approaches the dead cone threshold, the ratio increases both for ungroomed (upper half) and groomed jets (lower half), indicating that the enhancement due to additional logarithms $\log(\xi)$ in $\Sigma^b$ becomes significant. 
We note that for a given observable value $v \gtrsim \theta_D$, the ratio $\Sigma^b / \Sigma^q$  is the same for ungroomed and groomed jets. This can be understood from fixed order analysis.
Indeed, the additional contribution from grooming, the second term on the right hand side of Eq.~(\ref{eq:lambda-alpha-groomed-small-zcut}), is mass-independent, so it cancels out in the ratio.
However, for $v\ll\theta_D$, the Eq.~(\ref{eq:lambda-alpha-groomed-small-zcut} is no longer valid, and we obtain non-negligible differences between groomed and ungroomed results, see  Section 3.1 of Ref.~\cite{Dhani:2024gtx} for details.

To demonstrate the accuracy of our collinear approximation, we also plot the dashed black line, which represents the truncation of the ratio $\Sigma^b / \Sigma^q$ at $\Oas$.-
 The yellow line, on the other hand, represents the full real radiation matrix element at next-to-leading order (NLO) generated using \mgamcnlo~\cite{Alwall:2014hca}. For values of the observable $v \gtrsim \theta_D$, we find that our results approximate the full matrix element with an uncertainty of up to 10-20\%.

We now compare our NLL predictions with parton-level MC simulations from \pythia, \herwig, and \sherpa. 
We note that all our MC simulations are performed using their default settings, and at LO+PS accuracy.
The results are shown in Fig.~\ref{fig:LEP-all_pl} with solid red, green, and blue lines. 
For ungroomed observables, MC predictions differ significantly due to varying implementations of heavy particle radiation, see Refs.~\cite{Norrbin:2000uu, Bierlich:2022pfr}, \cite{Gieseke:2003rz, Hoang:2018zrp, Cormier:2018tog}, and \cite{Schumann:2007mg,Krauss:2016orf}, respectively. 
Notably, each MC overestimates the dead-cone effect, as $\Sigma^b / \Sigma^q$ increases before the dead cone boundary. 

The predictions for the groomed observables are given in the lower half of Fig.~\ref{fig:LEP-all_pl}. 
We used mMDT grooming with $\zcut = 0.1$. 
The NLL results for groomed $\Sigma^b / \Sigma^q$ ratios for $v<\theta_D$ show smaller enhancements compared to ungroomed ones. 
The grooming procedure also changes the slope and reduces discrepancies around the dead-cone threshold between NLL and MC predictions. 
Our NLL predictions for ECF and jet angularities differ at $v = \theta_D$. 
The $\Sigma^b / \Sigma^q$ for ECF is almost equal to unity, while jet angularities show a non-negligible deviation.
\begin{figure}
  \includegraphics[width=0.5\textwidth]{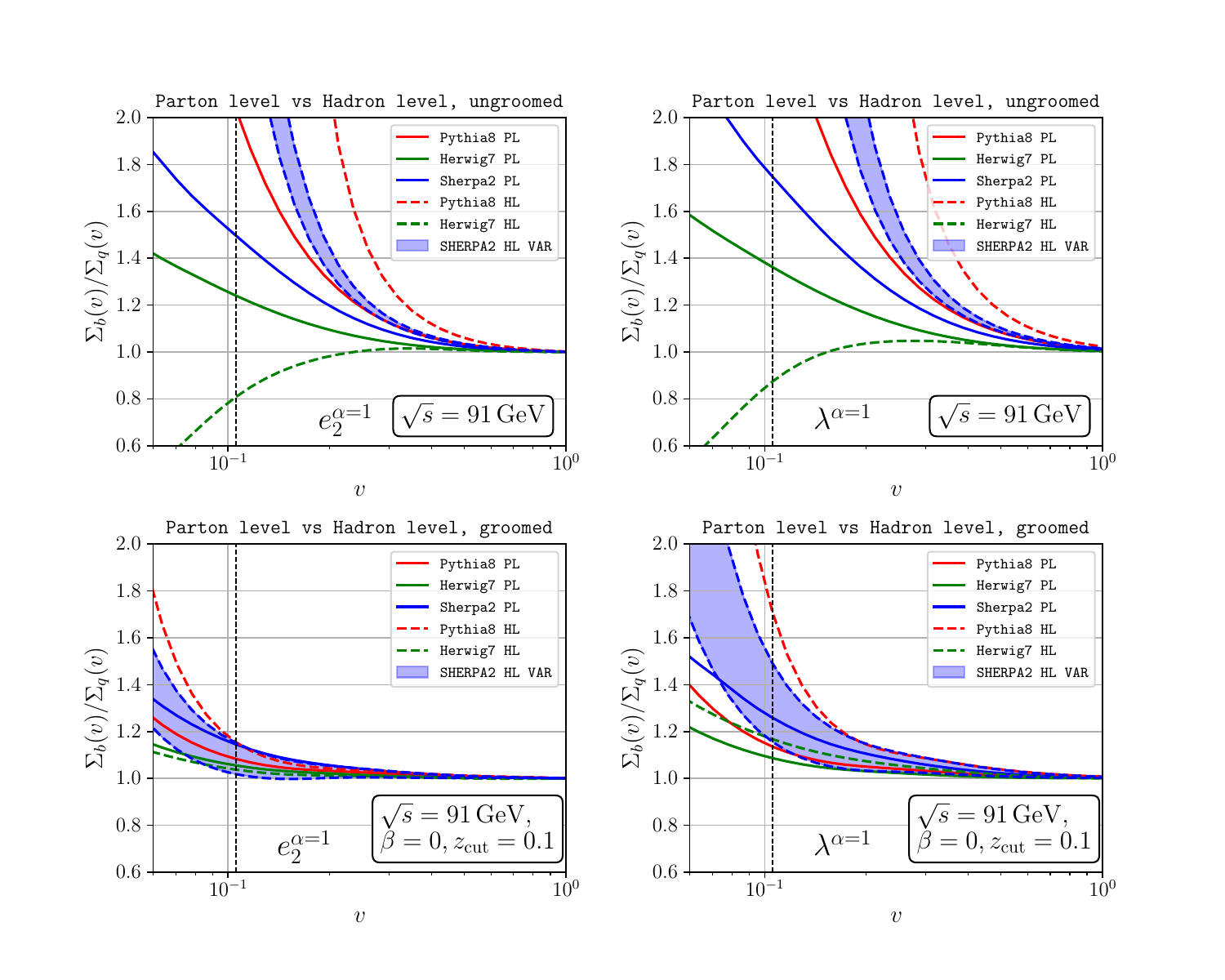}
  \caption{The ratio $\Sigma^b / \Sigma^q$ at parton (solid lines) and hadron (dashed lines) levels at LEP center-of-mass collision energy simulated with \pythia, \herwig and \sherpa MC is presented. The vertical dashed black line marks the dead-cone transition threshold $\theta_D = 2m_b / \sqrt{s} \approx 0.11$. The blue bands estimate the uncertainty due to two different hadronization models implemented in \sherpa.}
  \label{fig:LEP-all_hl}
\end{figure}

To estimate the impact of non-perturbative corrections on our results, we compare parton-level MC simulations (solid lines) with MC simulations that include hadronization corrections (dashed lines) in Fig.~\ref{fig:LEP-all_hl}. 
Currently, there are two primary models of hadronization available on the market: the Lund string model~\cite{Andersson:1983ia}, as implemented in \pythia, and the cluster model, as implemented in \herwig~\cite{Webber:1983if}. Unlike \pythia and \herwig, \sherpa incorporates both models with some modifications, as detailed in Ref.~\cite{Winter:2003tt}. Consequently, by altering the hadronization models in \sherpa, one can estimate the overall uncertainty associated with hadronization effects. This is illustrated in Fig.~\ref{fig:LEP-all_hl}, where the blue bands represent the uncertainty, with upper edge  corresponding to the Lund string model and the lower one to the cluster model\footnote{The rope hadronization model, developed by the \pythia collaboration~\cite{Bierlich:2017vhg}, in principle allows us to perform a similar estimate of hadronization uncertainty using \pythia. However, we note that this model is relatively new and has not yet been tuned to the data. Therefore, we refrain from using it in the current study.}.

Also, as noted in Ref.~\cite{Lee:2019lge}, the jet substructure of $b$-jets is influenced not only by non-perturbative corrections but also by $B$-hadron decays. 
In agreement with predictions of Ref.~\cite{Lee:2019lge} our previous  $pp$ study~\cite{Dhani:2024gtx} demonstrated that decay products of $B$-hadrons can significantly alter jet shapes. While such effects could be incorporated into our analysis using transfer matrices, as introduced in Ref.~\cite{Reichelt:2021svh}, we opt to simplify this study by treating all \mbox{$B$-hadrons} as stable in \pythia, \herwig, and \sherpa simulations. 
As a result, our findings can be directly compared to measurements where $B$-hadrons are reconstructed from their decay products.

Let us delve into the results shown in Fig.~\ref{fig:LEP-all_hl}. For the ungroomed case, hadronization significantly impacts the ratios $\Sigma^b / \Sigma^q$. Notably, the solid (parton-level) and dashed (hadron-level) curves for all three MC generators $-$ \pythia, \herwig, and \sherpa $-$ begin to diverge well before reaching the dead-cone threshold. 
Furthermore, these generators exhibit distinct predictions not only at the parton level (expected due to differences in their parton shower models) but also at the hadron level. 
The hadron-level discrepancies are usually smaller, as these MC generators have been tuned to provide consistent descriptions of existing data.
However, as evident in the figure, the \herwig yields predictions different from \pythia and \sherpa. Moreover, the \sherpa hadron-level uncertainty band does not overlap with its parton-level predictions. Consequently, we conclude that in the case of the ungroomed observable, the divergences between different event generators can highlight potential areas for improvement in current MC models and underscore the significance of analyzing archived LEP data to refine MC tunes further. It’s worth noting that such a pronounced influence of non-perturbative physics may effectively screen the essentially perturbative dead-cone effect for the setup considered in this paper.

Now, let’s focus on the groomed ratios presented in the lower half of Fig.~\ref{fig:LEP-all_hl}. The application of the mMDT groomer substantially reduces the discrepancies between parton-level and hadron-level results, particularly near the dead-cone threshold. Furthermore, the \sherpa hadronization uncertainty bands now overlap with parton-level \sherpa predictions, as well as with those of \pythia and \herwig. Notably, this effect is particularly pronounced for the case of the ECF observable, which exhibits reduced sensitivity to non-perturbative corrections around the dead-cone boundary compared to the jet angularity. This underscores the efficacy of grooming techniques in mitigating non-perturbative effects and enhancing the robustness of the analysis.

\begin{figure}
  \includegraphics[width=0.5\textwidth]{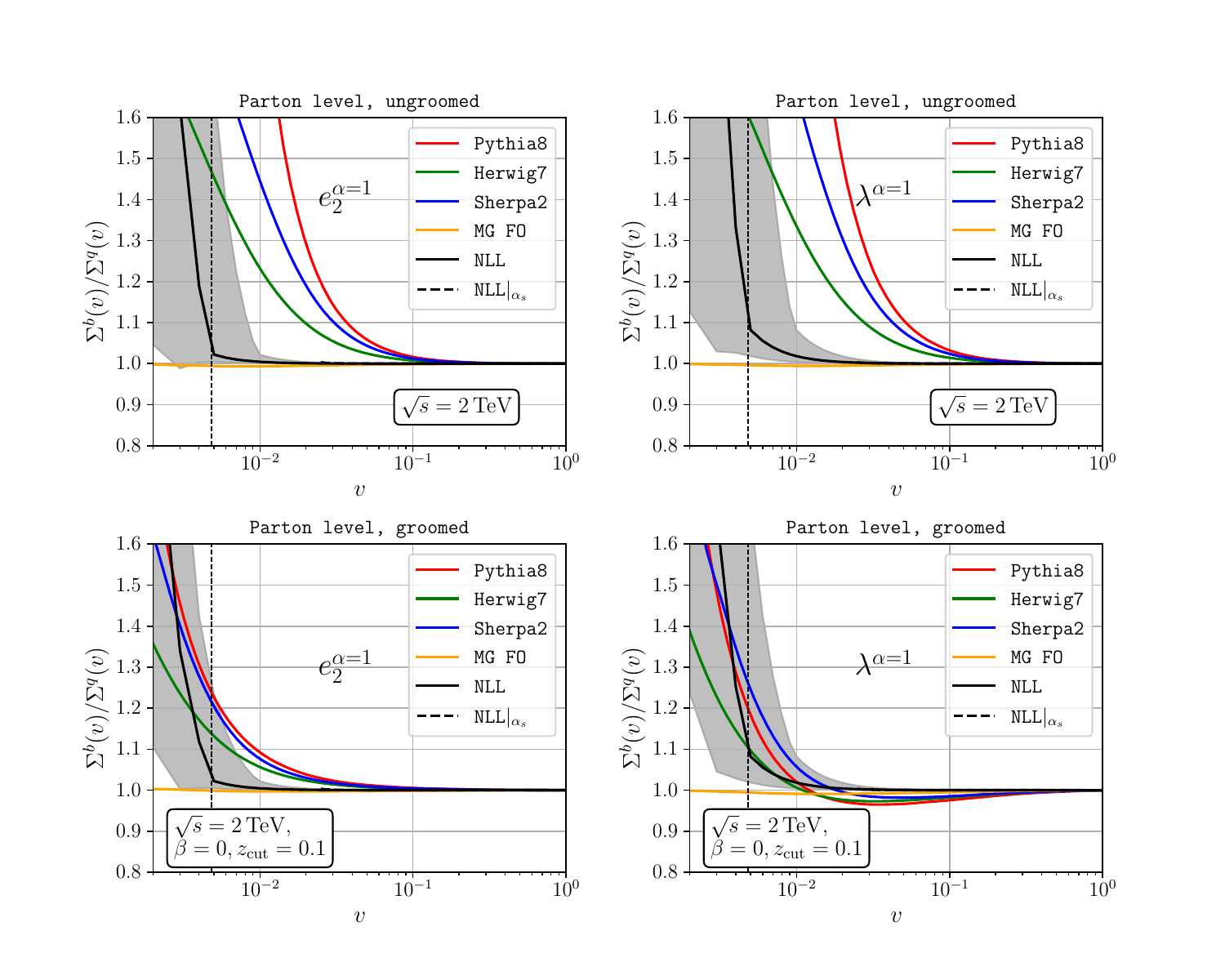}
  \caption{Same as in Fig.~\ref{fig:LEP-all_pl} but for $\sqrt{s} = 2$ TeV collision energy.}\label{fig:ILC-parton}
\end{figure}

We observe distinct behavior in the $\Sigma^b / \Sigma^q$ ratio simulated with \herwig between the ungroomed and groomed cases. Specifically, in the ungroomed scenario, the green dashed curve in Fig.~\ref{fig:LEP-all_hl} shows a decreasing trend at smaller observable values, differing from other predictions. This behavior arises from events containing only one particle per hemisphere, which results in trivial jet substructures and observable values of zero. These events are stacked in the first histogram bin (underflow bin) to account for their impact on normalization. Interestingly, we found that \herwig generates a significantly larger number of such events for the massless case compared to \pythia and \sherpa. However, in the case of groomed distributions, the population of the underflow bins between massive and massless cases changes and aligns with other predictions.
To conclude this section, let’s estimate the feasibility of observing the dead-cone effect with ECF and jet angularities at future $e^+ \, e^-$ colliders. In Fig.~\ref{fig:ILC-parton}, we compare our resummed predictions with MC simulations at \mbox{$\sqrt{s} = 2$ TeV} center-of-mass collision energy, which corresponds to the ILC setup. The increase in collision energy from 91 GeV to \mbox{2 TeV} implies that the ratios $\Sigma^b / \Sigma^q$ in the MC results start to deviate from unity at much smaller values, approximately at $v\sim 0.1$. Similarly, the dead-cone driven changes become visible if $v \lsim 0.01$, which is in agreement with the leading-log estimate of the dead-cone boundary. These small thresholds pose a significant challenge for future ILC measurements, as detecting the dead cone effect with jets seeded by $b$-quarks would require exceptional detector resolution. A potential solution to this problem could be studying multi-jet events instead of di-jet events, where the average energy per $b$-jet is smaller and hence increases the dead-cone threshold value. Additionally, one could also study the dead cone effect for top quarks, although it’s not yet clear if its extremely short lifetime would allow to produce enough QCD emissions to observe the suppression of collinear radiation. For now, we leave the study of multi-jet ILC events and top production to future work.

\section{Conclusions}
\label{sec:conclusions}
We explored jet substructure observables, focusing on the ECF, $e_2^{\alpha}$, and jet  angularity, $\lambda^{\alpha}$, measured on heavy-flavor jets in $e^+ \, e^-$ collisions. 
By calculating their cumulative distributions for both ungroomed and mMDT-groomed cases, we achieved NLL accuracy level, resumming logarithms of the observable, $\log v$, and  of the quark mass to hard scale ratio, $\log \xi$. 
While the two observables share an identical logarithmic structure, they diverge at terms beyond NLL. Partial fixed-order contributions were also incorporated to better capture the intricate dead-cone transition region.

To gain deeper insights, we compared the cumulative distributions for ungroomed and groomed jets relative to light-quark jets. Our predictions were benchmarked against results from \textsc{Pythia}, \textsc{Herwig}, and \textsc{Sherpa} MC simulations across various lepton collider energies, with special attention to the LEP energy of 
$\sqrt{s}=91$ GeV. 
Our findings reveal that for the analytical results and for $v\gtrsim \theta_D$, the ungroomed and groomed distributions align, but differ if $v\ll\theta_D$. 
The parton-level \textsc{Pythia}, \textsc{Herwig} and \textsc{Sherpa} MC  simulations demonstrate behaviour qualitatively similar to our NLL results, however, with significant quantitative differences. 
In particular, we found that all three different MC simulations show much stronger sensitivity to mass effects compared to our NLL calculations. 
Moreover, their predictions do not agree among themselves for sufficiently small observable values.
Even though \textsc{Pythia}, \textsc{Herwig}, and \textsc{Sherpa} are expected to demonstrate better agreement at the hadron level (since the non-perturbative models used in all these programs  were tuned to describe experimental data), we found that the parton-level differences do not go away after enabling the non-perturbative effects. 
Therefore, we conclude that future experimental studies of heavy-flavour jet substructure can be used to improve existing MC tunes and models of heavy-quark radiation.
We also found that the mMDT grooming can reduce the impact of non-perturbative effects and decrease discrepancies between the MC simulations and our NLL results. 
In this regard, we expect groomed versions of the jet substructure observables considered in this paper to be better candidates for testing resummed predictions, whereas ungroomed versions can be used to constraint non-perturbative models in general-purpose MC event generators.

We also extended our comparisons to the future ILC setup at a center-of-mass energy of $\sqrt{s}=2$ TeV. Here, the smaller dead-cone angle ($\theta_D \approx 0.005$) introduces significant challenges for detecting the effect in $b$-quark jets, emphasizing the need for exceptional detector resolution. Consequently, we propose that analyzing archived LEP data offers a more practical avenue for studying $b$-quark jet substructure than relying solely on future high-energy experiments. That said, exploring the dead-cone effect in multi-jet events and top-quark jets remains an exciting prospect for next-generation colliders like the ILC and \mbox{FCC~\cite{Maltoni:2016ays}.}

\begin{acknowledgments}
We express our gratitude to Simone Marzani and Gregory Soyez for  encouraging us to write this paper as well as for many fruitful discussions.
OF also would like to thank Yang-Ting Chien,  Max Knobbe, Simon Pl\"{a}tzer and Andreas Papaefstathiou  for useful and interesting discussions.
The work of PKD is supported by European Commission MSCA Action COLLINEAR-FRACTURE, Grant Agreement No. 101108573, and by the Spanish Government (Agencia Estatal de Investigaci\'on MCIN/AEI/10.13039/501100011033) Grants No.~PID2020-114473GB-I00, No. PID2023-146220NB-I00 and No. CEX2023-001292-S. 
The work of OF is supported in part by the US Department of Energy (DOE) Contract No.~DE-AC05-06OR23177, under which Jefferson Science Associates, LLC operates Jefferson Lab and by the Department of Energy Early Career Award grant DE-SC0023304.
The work of AG is supported by the Italian Ministry of Research (MUR), and by ICSC Spoke~2 under grant BOODINI.

Most of the simulation is conducted using the computing facilities of the Galileo cluster at the Department of Physics and Astronomy of Georgia State University. The generated events are analysed using the \rivet framework~\cite{Buckley:2010ar,Bierlich:2019rhm}. All plots in this paper are generated with the Matplotlib Python library~\cite{Hunter:2007ouj}. Our analytical predictions, MC run cards, and analysis files are available on request.
\end{acknowledgments}

\appendix
\section{$\Oas$ results for $\lambda^\alpha$ and $e_2^\alpha$}
\label{app: fixed order results}
Here, we report the fixed order expressions for $\mathcal{R}_{\lambda^\alpha}$ and $\mathcal{R}_{e_2^\alpha}$ computed in the quasi-collinear limit. We neglect power-corrections in $v$ and $\xi$, but retain the dependence on their ratio $x=\xi/\bar{v}^\frac{2}{\alpha}$. The result for $\mathcal{R}_{\lambda^\alpha}$  reads
\begin{align}
	\label{eq: Sudakov for lambda a}
	\mathcal{R}^{(\text{f.o.})}_{\lambda^\alpha}(v,\xi)
	&= \frac{\as\cf}{\pi}\Bigg[
	\frac{1}{\alpha}\log^2 \bar v
	-\frac{\alpha}{4}\log^2(1+x)
	+\frac{3}{2\alpha}\log \bar v \nonumber \\
	&+\left(\frac{3}{4}-\frac{\alpha}{2}\right)\log(1+x)
	+\frac{7}{4\alpha}-\frac{\alpha}{2}\text{Li}_2\left(\frac{x}{1+x}\right)\nonumber \\
	&+\frac{\alpha-2}{\alpha+2}x\,_2 F_1\left(1,1+\frac{\alpha}{2};2+\frac{\alpha}{2};-x\right)\nonumber \\
	&+\frac{x}{4(\alpha+1)}\,_2 F_1\left(1,1+\alpha;2+\alpha;-x\right)\Bigg],
	\end{align}
and the radiator associated with the observable $e_2^\alpha$ is given by
	\begin{align}\label{eq: Sudakov for ECF}
		&\mathcal{R}^{(\text{f.o.})}_{e_2^{\alpha}}(v,\xi)=\mathcal{R}^{(\text{f.o.})}_{\lambda^\alpha}(v,\xi)
		+\frac{\as \cf}{\pi}
		\nn\\
		&\times\Bigg\{\int^1_0 \de z \frac{1+(1-z)^2}{2z}\log\left(\frac{(1-z)^\frac{2}{\alpha}+((1-z)z)^\frac{2}{\alpha}x}{1+((1-z)z)^\frac{2}{\alpha}x}\right)
		\nn\\
		&+\int^1_0 \de z~
		\frac{1-z}{z}\left[\frac{z^{\frac{2}{\alpha}}x}{1+z^\frac{2}{\alpha}x}-\frac{((1-z)z)^{\frac{2}{\alpha}}x}{1+((1-z)z)^\frac{2}{\alpha}x}\right] \Bigg\}.
	\end{align}
The calculation of the ECF $e_2^\alpha$ is more complex due to the presence of products $z_i z_j$, as can be seen in Eq.~(\ref{eq:e2}). These terms prevent a closed-form analytic expression for a generic value of $\alpha$.
\section{All-order results for light-quark jets}
\label{sec:resummed-radiators-light-quarks}
We briefly summarize the resummed results for the cumulative ditributions for the light quarks. 
The ungroomed cumulative distribution reads as
\begin{align}
\label{eq:sigma_q_w_corr}
\Sigma_V^q(v,\xi)=& \frac{e^{-2\gamma_{\text{E}} R_q'(v,\xi)}}{\Gamma\left(1+ 2R_q'(v,\xi)\right)} e^{-2R_q(v,\xi)} \nonumber \\ 
&\times \exp \left[-2\left( \mathcal{R}^{(\text{f.o.})}_V(v,0)- R_q^\text{(f.c.)}(v,0)\right)\right].
\end{align}
At NLL accuracy, the radiator for all observables is found to be
\begin{align}\label{eq: massless R(v)}
R_q(v,\xi)=\int^{1}_{0} \frac{\de \theta^2}{\theta^2}\int^{1}_0 \de z P_{gq}\left(z\right)\frac{\as^{\text{CMW}}(\bar k_t^2)}{2\pi} \Theta\left(z \theta^\alpha-\bar v\right),
\end{align}
where, $P_{gq}(z)$ is the standard (massless) splitting function
\begin{equation}
P_{gq}(z)= \cf \frac{1+(1-z)^2}{z}.
\end{equation}
Note that $\Sigma_V^q(v,\xi)$ retains a dependence on $\xi$ because the integral over the running coupling is performed in the decoupling scheme, see Eqs.~(\ref{eq: dec_a}) and (\ref{eq: dec_b}).
The groomed cumulative distribution is built in the same way and it is given by
\begin{align}
\label{eq:sigma_q_w_corr_SD}
\bar\Sigma_V^q(v,\xi)=&  \frac{e^{-2\gamma_{\text{E}} \bar R_q'(v,\xi)}}{\Gamma\left(1+ 2\bar R_q'(v,\xi)\right)} e^{-2\bar R_q(v,\xi)} \nonumber \\
&\times \exp \left[-2\left( \mathcal{R}^{(\text{f.o.})}_V(v,0)- R_q^\text{(f.c.)}(v,0)\right)\right],
\end{align}
where the associated radiator is
\begin{align}\label{eq: massless R(v) SD}
\bar{R}_q(v,\xi)=\int^{1}_{0} \frac{\de \theta^2}{\theta^2}\int^{1}_{\zc} \de z P_{gq}\left(z\right)\frac{\as^{\text{CMW}}(\bar k_t^2)}{2\pi} \Theta\left(z \theta^\alpha-\bar v\right).
\end{align}

\bibliographystyle{jhep}
\bibliography{references}
\end{document}